\documentclass[fleqn,twoside]{article}
\usepackage{espcrc2}
\usepackage{epsfig}
\usepackage{rotate}

\newcommand{\AmS}{{\protect\the\textfont2
  A\kern-.1667em\lower.5ex\hbox{M}\kern-.125emS}}

\title{
\vspace{-1.8cm}
\hfill \rm \null \hfill
 \hbox{\normalsize ADP-01-56/T488} \\
\vspace{+1.3cm}
Quenched chiral perturbation theory for baryon form factors}

\author{Derek B. Leinweber
\address{Special Research Centre for the Subatomic Structure of
Matter, and Department of Physics and Mathematical Physics, University
of Adelaide, SA, 5005, Australia}}%
       
\begin{document}
\thispagestyle{empty}

\begin{abstract}
A new intuitive method for the rapid calculation of the leading
nonanalytic behavior of hadronic observables in quenched chiral
perturbation theory is presented.  After proving the technique in a
consideration of baryon masses, the quenched magnetic moments of octet
baryons are addressed.  The technique provides a separation of
magnetic moment contributions into full, sea, valence and quenched
valence contributions, the latter being the conventional view of the
quenched approximation.  Both baryon mass and meson mass violations of
SU(3)-flavor symmetry are accounted for.  A comprehensive examination
of the individual quark-sector contributions to octet baryon magnetic
moments reveals that the $u$-quark contribution to the proton magnetic
moment provides an optimal opportunity to directly view nonanalytic
behavior associated with the quenched meson cloud of baryons in the
quenched approximation.
\vspace{1pc}
\end{abstract}

\maketitle

\section{INTRODUCTION}

Separation of the valence and sea-quark-loop contributions to the
meson cloud of full QCD hadrons is a non-trivial task.  Early
calculations addressing the meson cloud of mesons employed a
diagrammatic method \cite{Sharpe:1992ft}.  The formal theory of quenched
chiral perturbation theory (Q$\chi$PT) was subsequently established in
Ref.~\cite{Bernard:1992mk}.  There, meson properties were examined in a
formulation where extra commuting ghost-quark fields are introduced to
eliminate the dependence of the path integral on the fermion-matrix
determinant.  This approach was extended to the baryon sector in
Ref.~\cite{Labrenz:1996jy}.

While the formalism of Q$\chi$PT is essential to establishing the
field theoretic properties, it is desirable to formulate an efficient
and perhaps more intuitive approach to the calculation of quenched
chiral coefficients.  Rather than introduce extra degrees of freedom
to remove the effects of sea-quark loops, the approach described
herein describes the systematic separation of valence and
sea-quark-loop contributions.  Upon removing the contributions of
sea-quark-loops, one arrives at the conventional view of quenched
chiral perturbation theory.

Since the presentation of this talk, there has been a resurgence in
Q$\chi$PT calculations.  In particular, the magnetic moments of octet
baryons have been examined \cite{Savage:2001dy} using the formal approach of
Q$\chi$PT \cite{Labrenz:1996jy}.  There the leading-nonanalytic (LNA)
behavior of the magnetic moment for each baryon of the octet is
calculated.  The formal approach completely eliminates all
sea-quark-loop contributions to baryon moments.

However, sea-quark loops do make a contribution to matrix elements in
the quenched approximation.  Insertion of the current in calculating
the three-point correlation function provides pair(s) of
quark-creation and annihilation operators.  These can be contracted
with the quark field operators of the hadron interpolating fields
providing ``connected insertions'' of the current, or self-contracted
to form a sea-quark-loop contribution or ``disconnected insertion'' of
the current.  The latter contributions to baryon electromagnetic form
factors have already been examined in quenched simulations
\cite{Dong:1997xr,Wilcox:2000qa}.  Hence in formulating quenched chiral perturbation
theory it is important to provide an opportunity to include these
particular sea-quark-loop contributions.

In the following, a new intuitive method is presented for the rapid
calculation of the quenched chiral coefficients of the LNA terms of
QCD.  While the quenched $\eta'$ also gives rise to new nonanalytic
behavior, such effects on baryon magnetic moments are numerically
small in the region of interest \cite{Savage:2001dy} and are not
addressed further in this brief report.  Sec.~\ref{sec:mass} proves
the technique via a consideration of baryon masses.  The derivation of
the quenched chiral coefficients for the quenched magnetic moments of
octet baryons is described in Sec.~\ref{sec:mom}.  The technique
provides a separation of magnetic moment contributions into full, sea,
valence and quenched valence contributions, the latter being the
conventional view of the quenched approximation.  Sec.~\ref{sec:mom}
also outlines how both baryon mass and meson mass violations of
SU(3)-flavor symmetry are accounted for.  A comprehensive examination
of the individual quark-sector contributions to octet baryon magnetic
moments is presented in Sec.~\ref{sec:results}.

\section{QUENCHED BARYON MASSES}
\label{sec:mass}

\begin{figure*}[t]
\begin{center}
\setlength{\unitlength}{1.0cm}
\setlength{\fboxsep}{0cm}
\begin{picture}(15,4.2)
\put(0,0){\begin{picture}(3,4)\put(0,0){
\epsfig{file=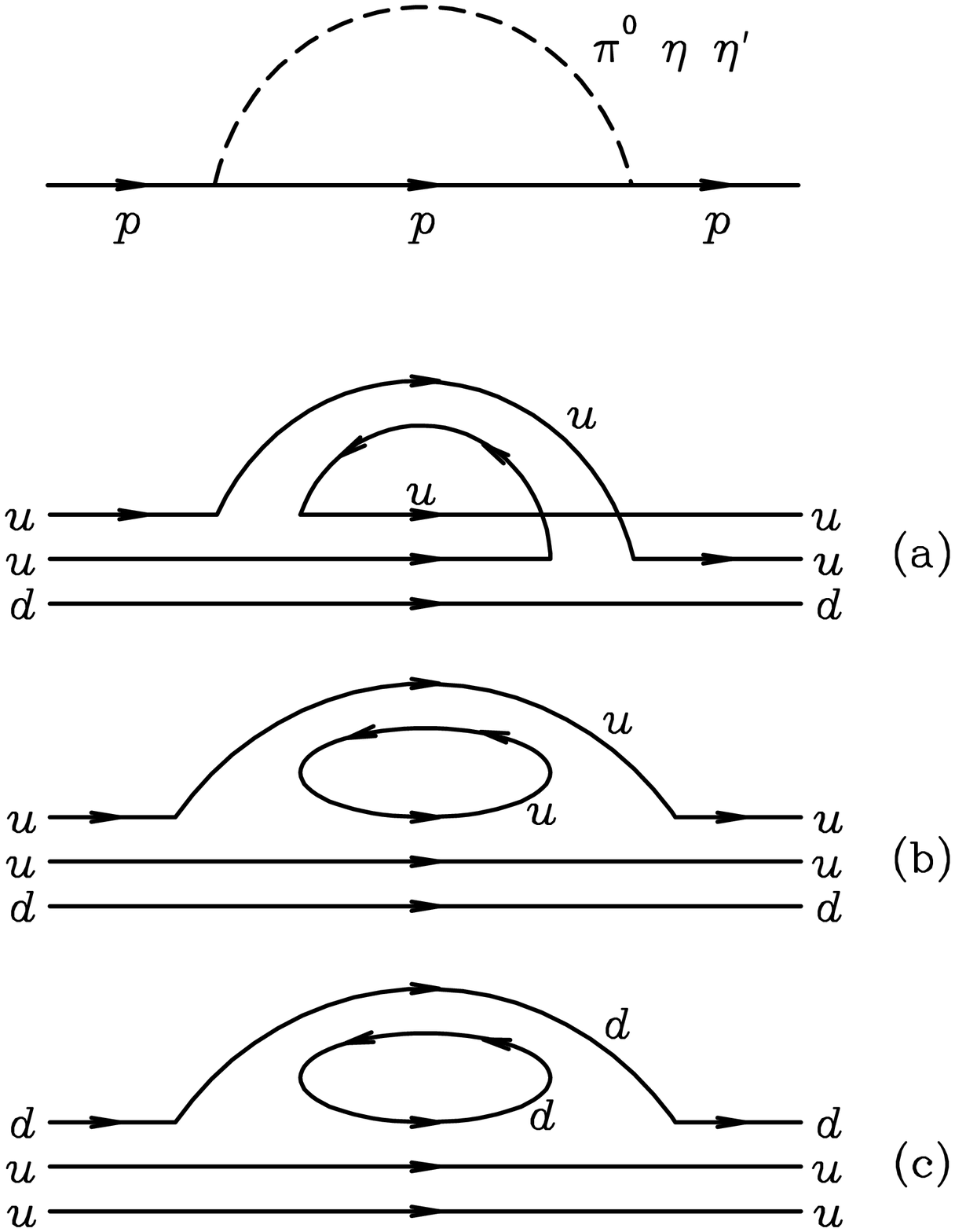,width=3.0cm}}\end{picture}}
\put(4.0,0){\begin{picture}(3,4)\put(0,0){
\epsfig{file=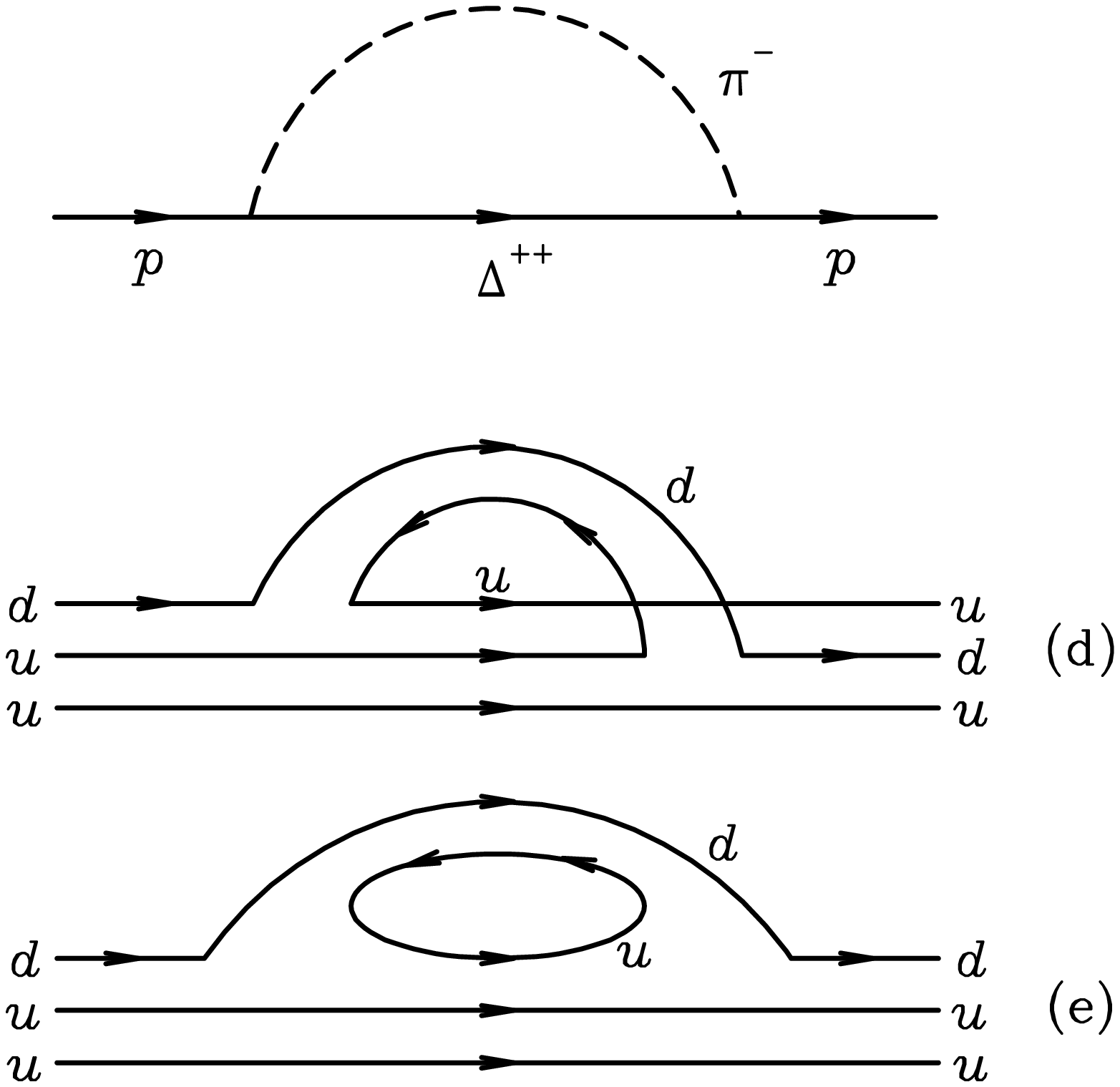,width=3.0cm}}\end{picture}}
\put(8.0,0){\begin{picture}(3,4)\put(0,0){
\epsfig{file=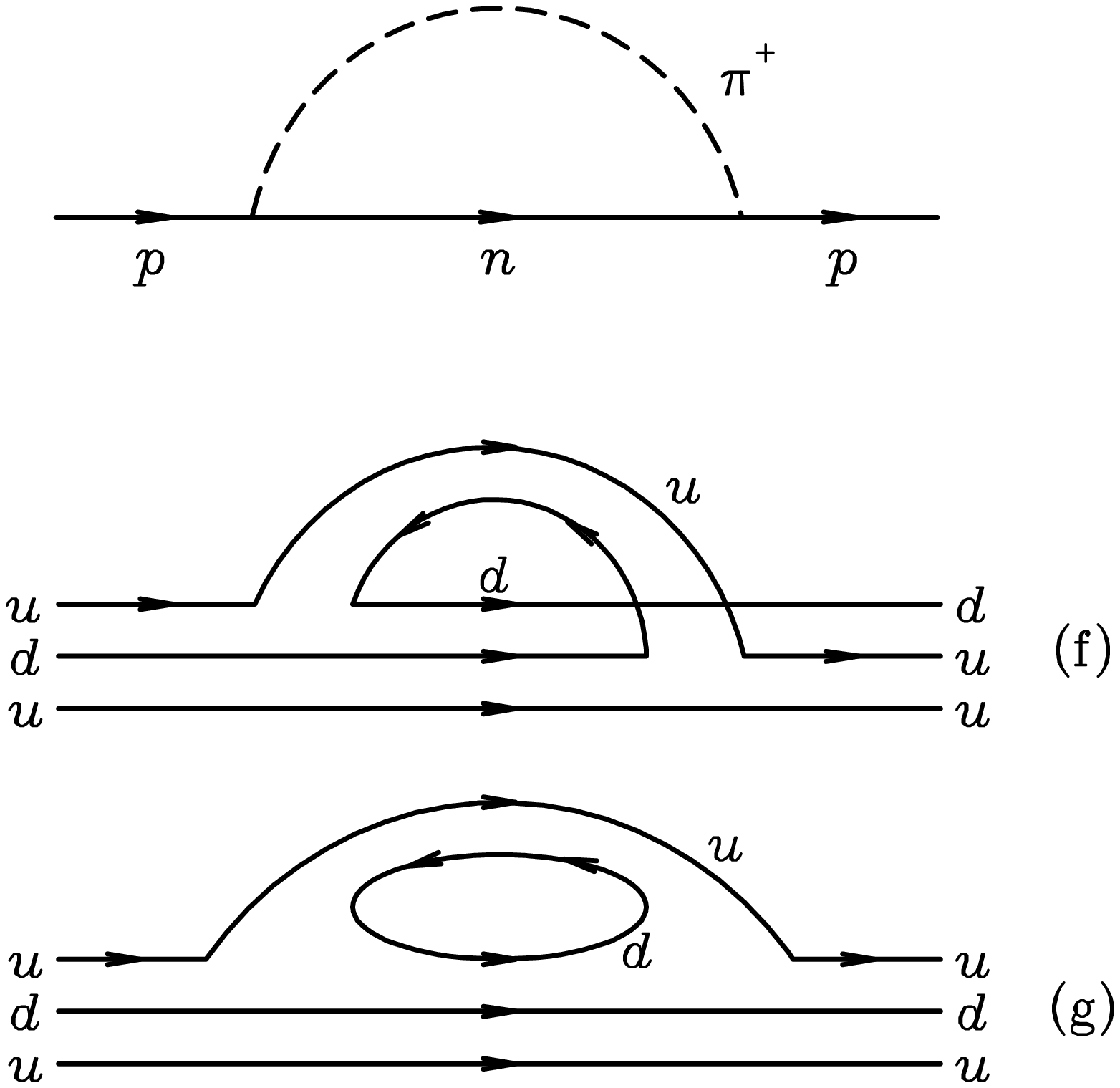,width=3.0cm}}\end{picture}}
\put(12,0){\begin{picture}(3,4)\put(0,0){
\epsfig{file=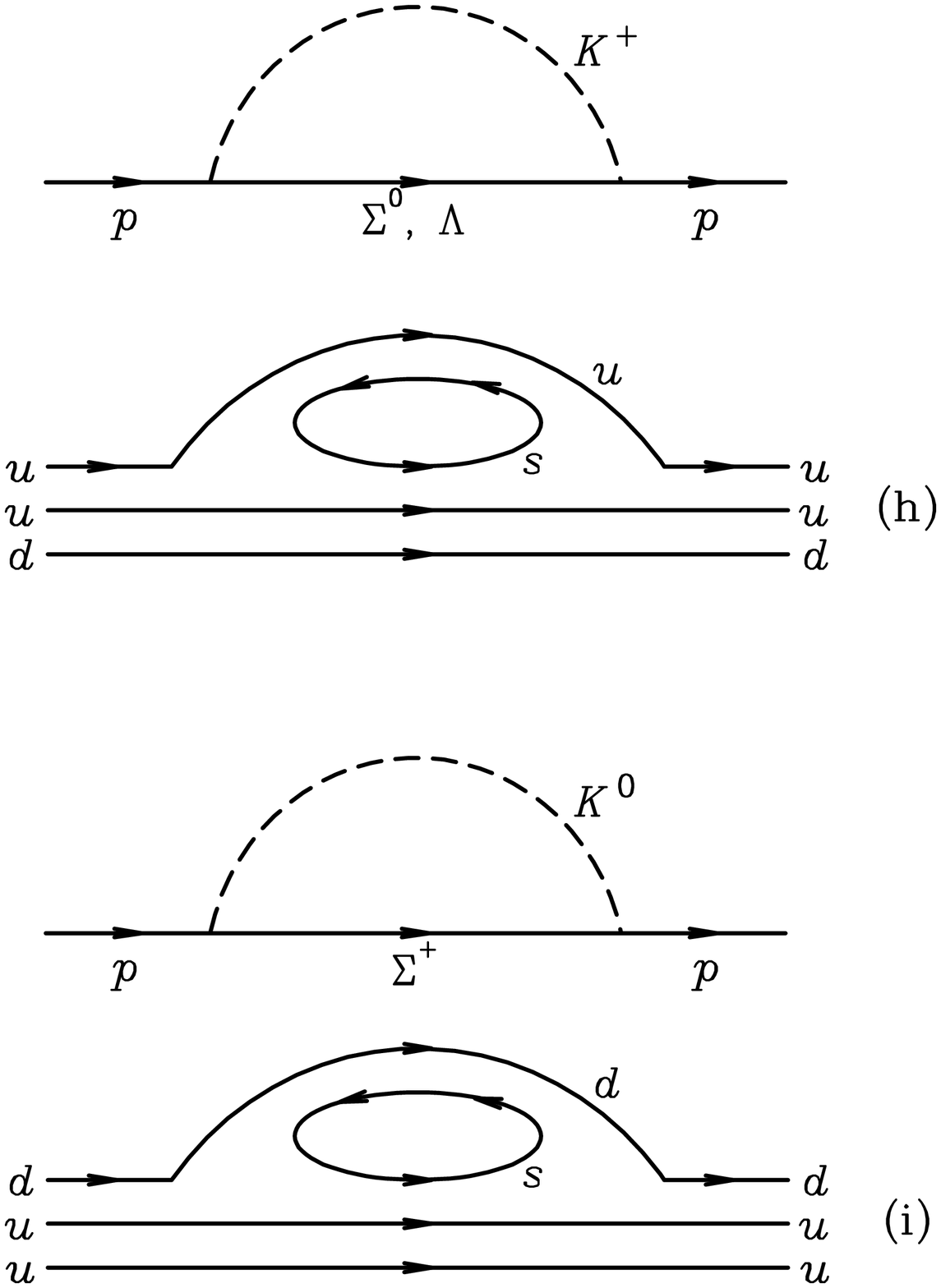,  width=3.0cm}}\end{picture}}
\end{picture}
\end{center}
\vspace{-1.0cm}
\caption{The pseudo-Goldstone meson cloud of the proton and the
associated quark flow diagrams.}
\label{ProtonCloud}
\end{figure*}

The following calculations are simplified through the use of the
standard octet and singlet interaction Lagrangians
\cite{Rijken:1998yy} in which meson-baryon couplings are subsequently
expressed in terms of the $F$ and $D$ coupling coefficients.  For
example, $f_{NN\pi} = F + D$, $f_{\Sigma N K} = D - F$, etc.  In the
following, numerical estimates are based on the one-loop corrected
values \cite{Jenkins:1992pi} of $F = 0.40$ and $D = 0.61$, with $f_\pi
= f_K = 93$ MeV.

The quark flow diagrams of Fig.~\ref{ProtonCloud} illustrate the
processes which give rise to the LNA behavior of proton observables.
Table~\ref{tab:massCloud} summarizes the contributions of the $\pi$-,
$\eta$- and $\eta'$-cloud diagrams of Fig.~\ref{ProtonCloud} labeled
by the corresponding quark-flow diagrams.  The suppression of
sea-quark-loops in the quenched approximation renders the $\eta'$ mass
degenerate with the pion.  The octet-decuplet transition couplings are
taken to be negligible.  Summation of these couplings and
incorporation of the factors from the loop integral provides the LNA
term
\[
- ( 3 F^2 + D^2 ) \, {m_\pi^3 \over 8 \pi  f_\pi^2} 
= -3.92\, m_\pi^3 \, .
\]

\begin{table}[t]
\caption{Meson-cloud contributions of Fig.~\protect\ref{ProtonCloud}.}
\label{tab:massCloud}
\begin{tabular}{cccc}
\hline
\noalign{\smallskip}
Fig.    &Channel     &Coupling           &Coupling \\
\hline 		     
\hline		     
\noalign{\smallskip}
a,b,c   &$p\, \pi^0$ &$f_{NN\pi}^2$      &$(F+D)^2$ \\
a,b,c   &$p\, \eta$  &$f_{NN\eta}^2$     &$(3F-D)^2/3$ \\
a,b,c   &$p\, \eta'$ &$f_{NN\eta'}^2$    &$2(3F-D)^2/3$ \\
f,g     &$n\, \pi^+$ &$2f_{NN\pi}^2$     &$2(F+D)^2$ \\
\hline
\end{tabular}
\end{table}

\begin{table}[t]
\caption{Sea-quark-loop contributions of Fig.~\protect\ref{ProtonCloud}.}
\label{tab:massLoop}
\begin{tabular}{cccc}
\hline
\noalign{\smallskip}
Fig.    &Channel          &Coupling           &Coupling \\
\hline 			  
\hline			  
\noalign{\smallskip}
b       &$\Lambda\,  K^+$ &$f_{\Lambda N K}^2$      &$(3F+D)^2/3$ \\
b       &$\Sigma^0\, K^+$ &$f_{\Sigma N K}^2$       &$(D-F)^2$ \\
c       &$\Sigma^+\, K^0$ &$2f_{\Sigma N K}^2$      &$2(D-F)^2$ \\
e       &$\Sigma^+\, K^0$ &$2f_{\Sigma N K}^2$      &$2(D-F)^2$ \\
g       &$\Lambda\,  K^+$ &$f_{\Lambda N K}^2$      &$(3F+D)^2/3$ \\
g       &$\Sigma^0\, K^+$ &$f_{\Sigma N K}^2$       &$(D-F)^2$ \\
\hline			  
\end{tabular}
\end{table}

The separation of the meson cloud into valence and sea-quark
contributions is shown in the quark-flow diagrams labeled by letters.
To quench the theory, one must understand the chiral behavior of the
valence-quark loops of Figs.~\ref{ProtonCloud}(a), (d) and (f) and the
sea-quark loops of Figs.~\ref{ProtonCloud}(b), (c), (e), (g), (h) and
(i) separately.  If one can isolate the behavior of the diagrams
involving a quark loop, then one can use the known LNA behavior of the
full meson-based diagrams to extract the corresponding valence-loop
contributions.

For example, Fig.~\ref{ProtonCloud}(b) involves a $u$-quark loop where
no exchange term is possible.  Thus the $u$-quark in the loop is
distinguishable from all the other quarks in the diagram.  The chiral
structure of this diagram is therefore identical to that for a
``strange'' quark loop, as illustrated in Fig.\ \ref{ProtonCloud}(h),
provided the ``strange'' quark in {\it this} case is understood to
have the same mass as the $u$-quark.

The corresponding hadron diagram which gives rise to the LNA structure
of Fig.\ \ref{ProtonCloud}(b) is therefore the $K^+$-loop diagram
above Fig.~\ref{ProtonCloud}(h), with the distinguishable ``strange''
quark mass set equal to the mass of the $u$-quark.  That is, the
intermediate baryon masses appearing in the $K^+$-loop diagram are
degenerate with the nucleon.  Similarly, the ``kaon'' mass is
degenerate with the pion.

The sum of the first two lines of Table~\ref{tab:massLoop} provides
the contribution of diagram Fig.~\ref{ProtonCloud}(b).  Similar
arguments allow one to establish the remaining loop contributions to
the light-meson cloud.  Summing the couplings of
Table~\ref{tab:massLoop} indicates sea-quark-loops contribute
\[
- ( 9 F^2 - 6 F D + 5 D^2 ) \, {m_\pi^3 \over 24 \pi
f_\pi^2} 
= -2.82\, m_\pi^3 \, ,
\]
to the LNA behavior of the nucleon such that the
net quenched contribution is
\[
- (3 F D - D^2 ) \, {m_\pi^3 \over 12 \pi  f_\pi^2} 
= -1.10\, m_\pi^3 \, ,
\]
in agreement with the more formal approach of Labrenz and Sharpe
\cite{Labrenz:1996jy}.

\section{BARYON MAGNETIC MOMENTS}
\label{sec:mom}

\begin{table*}[t]
\caption{Determination of the $u$-quark contribution to the proton
magnetic moment as illustrated in Fig.~\protect\ref{ProtonCloud}.}
\label{tab:mag}
\begin{tabular}{cccclcc}
\\
\multicolumn{7}{l}{\bf Total Contributions} \\
\hline
Diagram   &Channel         &Mass        &Charge &Term                                      &$\beta$       &$\chi$  \\
\hline 
\hline
f,g     &$n\, \pi^+$      &$N \pi$     &$+1$   &$+2 f_{NN\pi}^2       \, m_\pi$           &$-(F+D)^2$    &$-4.41$ \\
h       &$\Sigma^0\, K^+$ &$\Sigma K$  &$+1$   &$+  f_{\Sigma N K}^2  \, m_{N \Sigma K}$  &$-(D-F)^2/2$  &$-0.10$ \\
h       &$\Lambda \, K^+$ &$\Lambda K$ &$+1$   &$+  f_{\Lambda N K}^2 \, m_{N \Lambda K}$ &$-(3F+D)^2/6$ &$-2.36$ \\
\noalign{\smallskip}
\hline
\\
\multicolumn{7}{l}{\bf Sea-Quark Loop Contributions} \\
\hline
Diagram   &Channel         &Mass     &Charge   &Term                            &$\beta$    &$\chi$  \\
\hline 
\hline
b       &$\Lambda\,  K^+$ &$N \pi$  &$-1$     &$-f_{\Lambda N K}^2 \, m_\pi$   &$(3F+D)^2/6$ &$+2.36$ \\
b       &$\Sigma^0\, K^+$ &$N \pi$  &$-1$     &$-f_{\Sigma N K}^2  \, m_\pi$   &$(D-F)^2/2$  &$+0.10$ \\
e       &$\Sigma^+\, K^0$ &$N \pi$  &$-1$     &$-2f_{\Sigma N K}^2 \, m_\pi$   &$(D-F)^2$    &$+0.19$ \\
Total   &                 &         &         &                                &             &$+2.65$ \\
\hline
\multicolumn{7}{l}
{{\bf Net Valence Contribution:}\quad $-7.06\, m_\pi -0.10\, m_{N \Sigma K} -2.36\, m_{N \Lambda K}$} \\
\\
\multicolumn{7}{l}{\bf Quenching Considerations} \\
\hline
Diagram   &Channel         &Mass     &Charge   &Term                            &$\beta$       &$\chi$  \\
\hline 
\hline
b       &$\Lambda\,  K^+$ &$N \pi$  &$+1$     &$+f_{\Lambda N K}^2 \, m_\pi$   &$-(3F+D)^2/6$ &$-2.36$ \\
b       &$\Sigma^0\, K^+$ &$N \pi$  &$+1$     &$+f_{\Sigma N K}^2  \, m_\pi$   &$-(D-F)^2/2$  &$-0.10$ \\
g       &$\Lambda\,  K^+$ &$N \pi$  &$+1$     &$+f_{\Lambda N K}^2 \, m_\pi$   &$-(3F+D)^2/6$ &$-2.36$ \\
g       &$\Sigma^0\, K^+$ &$N \pi$  &$+1$     &$+f_{\Sigma N K}^2  \, m_\pi$   &$-(D-F)^2/2$  &$-0.10$ \\
h       &$\Sigma^0\, K^+$ &$\Sigma K$  &$+1$   &$+  f_{\Sigma N K}^2  \, m_{N \Sigma K}$  &$-(D-F)^2/2$  &$-0.10$ \\
h       &$\Lambda \, K^+$ &$\Lambda K$ &$+1$   &$+  f_{\Lambda N K}^2 \, m_{N \Lambda K}$ &$-(3F+D)^2/6$ &$-2.36$ \\
\noalign{\smallskip}
\hline
\multicolumn{7}{l}
{{\bf Net Quenched Valence Contribution:}\quad $-2.15 \, m_\pi$}
\end{tabular}
\end{table*}

The LNA contribution to baryon magnetic moments proportional to
$m_\pi$ or $m_K$ has its origin in couplings of the electromagnetic
(EM) current to the meson propagating in the intermediate meson-baryon
state.  In order to pick out a particular quark-flavour contribution,
one sets the electric charge for the quark of interest to one and the
charge of all other flavours to zero.

Table~\ref{tab:mag} reports results for the $u$-quark in the proton.
The total contributions are calculated in the standard way, but with
charge assignments for the intermediate mesons (indicated in the
Charge column) reflecting in this case $q_u = 1$ and $q_d = q_s = 0$.
The extra baryon subscripts on the meson masses are a reminder of the
baryons participating in the diagram to facilitate more sophisticated
treatments of the loop integral in which baryon mass splittings are
taken into account.  The LNA contribution is $\beta \,
{m_N \over 8 \pi f_\pi^2 } \, m_\pi \equiv \chi \, m_\pi$ with $\beta$
and $\chi$ indicated in the last two columns.

``Sea-quark-loop contributions'' are contributions in which the EM
current couples to a sea-quark loop, in this case a $u$ quark.  Using
the techniques described in Sec.~\ref{sec:mass}, one can calculate the
contributions of these loops alone to the baryon magnetic moment.  The
Mass column of Table~\ref{tab:mag} is a reminder that the mass of the
``kaon'' considered in determining the coupling is actually the pion
mass for diagrams \ref{ProtonCloud}(b) and (e).  These diagrams will
contribute, even in the quenched approximation, when disconnected
insertions of the EM current are included in 
simulations \cite{Dong:1997xr,Wilcox:2000qa}.

Subtracting the sea-quark-loop contributions from the total
contributions provides the valence contribution of full QCD.
``Quenching considerations'' focus on diagrams in which the EM current
couples to a valence quark in a meson composed with a sea-quark loop.
Subtracting off these couplings from the valence contribution provides
the quenched valence contribution.

\section{RESULTS}
\label{sec:results}

\begin{table}[t]
\caption{Coefficients, $\chi$, providing the LNA contribution to
baryon magnetic moments by quark sectors normalized to unit charge.
Intermediate (Int.) meson-baryon channels are indicated to allow for
SU(3)-flavour breaking in both the meson and baryon masses.  Total,
sea-quark loop, valence (Val.) and quenched valence (Qch.)
coefficients are indicated.  Note $\epsilon = 0.0004$. }
\label{tab:quarkMom}
\addtolength{\tabcolsep}{-1pt}
\begin{tabular}{lccccc}
\hline
$q$   &Int.           &Total     &Loop     &Val.     &Qch. \\
\hline 
\hline 
\noalign{\smallskip}
$u_p$ &$N \pi$        &$- 4.41$  &$+ 2.65$ &$- 7.06$ &$- 2.15$ \\     
      &$\Lambda K$    &$- 2.36$  &$     0$ &$- 2.36$ &$     0$ \\     
      &$\Sigma  K$    &$- 0.10$  &$     0$ &$- 0.10$ &$     0$ \\     
\noalign{\smallskip}
$d_p$ &$N \pi$        &$+ 4.41$  &$+ 2.65$ &$+ 1.76$ &$+ 2.15$ \\     
      &$\Sigma  K$    &$- 0.19$  &$     0$ &$- 0.19$ &$     0$ \\     
\noalign{\smallskip}
$s_p$ &$\Lambda K$    &$+ 2.36$  &$+ 2.36$ &$     0$ &$     0$ \\     
      &$\Sigma K$     &$+ 0.29$  &$+ 0.29$ &$     0$ &$     0$ \\     
\noalign{\smallskip}
\hline 
\noalign{\smallskip}
$u_{\Sigma^+}$   &$\Sigma  \pi$  &$- 1.38$  &$+ 1.38$ &$- 2.77$ &$     0$ \\     
                 &$\Lambda \pi$  &$- 1.07$  &$+ 1.07$ &$- 2.15$ &$     0$ \\     
                 &$N K        $  &$     0$  &$+ 0.19$ &$- 0.19$ &$- 0.19$ \\     
                 &$\Xi K      $  &$- 4.41$  &$     0$ &$- 4.41$ &$- 1.95$ \\     
\noalign{\smallskip}
$d_{\Sigma^+}$   &$\Sigma  \pi$  &$+ 1.38$  &$+ 1.38$ &$     0$ &$     0$ \\     
                 &$\Lambda \pi$  &$+ 1.07$  &$+ 1.07$ &$     0$ &$     0$ \\     
                 &$N K        $  &$+ 0.19$  &$+ 0.19$ &$     0$ &$     0$ \\     
\noalign{\smallskip}
$s_{\Sigma^+}$   &$N K        $  &$- 0.19$  &$     0$ &$- 0.19$ &$+ 0.19$ \\     
                 &$\Xi K      $  &$+ 4.41$  &$+ 2.46$ &$+ 1.95$ &$+ 1.95$ \\     
\noalign{\smallskip}
\hline 
\noalign{\smallskip}
$u_\Lambda|d_\Lambda$
              &$\Sigma \pi $  &$+ 0.00$  &$+ 1.07$   &$- 1.07$   &$+ 0.00$ \\     
              &$\Lambda\eta_l$&$+ 0.00$  &$\epsilon$ &$-\epsilon$ &$+ 0.00$ \\     
              &$N K        $  &$+ 2.36$  &$+ 1.57$   &$+ 0.79$   &$+ 0.79$ \\     
              &$\Xi K      $  &$- 0.25$  &$+ 0.00$   &$- 0.25$   &$+ 0.29$ \\     
\noalign{\smallskip}
$s_\Lambda$   &$\Lambda\eta_s$&$+ 0.00$  &$+ 1.57$   &$- 1.57$   &$+ 0.00$ \\     
              &$N K        $  &$- 4.72$  &$+ 0.00$   &$- 4.72$   &$- 1.57$ \\     
              &$\Xi K      $  &$+ 0.50$  &$+ 1.07$   &$- 0.57$   &$- 0.57$ \\     
\noalign{\smallskip}
\hline
\noalign{\smallskip}
$u_{\Xi^0}$   &$\Xi \pi$      &$- 0.19$  &$+ 0.19$ &$- 0.38$ &$     0$ \\     
              &$\Lambda K$    &$     0$  &$+ 0.25$ &$- 0.25$ &$- 0.25$ \\     
              &$\Sigma K$     &$+ 4.41$  &$+ 2.21$ &$+ 2.21$ &$+ 2.21$ \\     
              &$\Omega K$     &$     0$  &$     0$ &$     0$ &$+ 0.19$ \\     
\noalign{\smallskip}
$d_{\Xi^0}$   &$\Xi \pi$      &$+ 0.19$  &$+ 0.19$ &$     0$ &$     0$ \\     
              &$\Lambda K$    &$+ 0.25$  &$+ 0.25$ &$     0$ &$     0$ \\     
              &$\Sigma K$     &$+ 2.21$  &$+ 2.21$ &$     0$ &$     0$ \\     
\noalign{\smallskip}
$s_{\Xi^0}$   &$\Lambda K$    &$- 0.25$  &$     0$ &$- 0.25$ &$+ 0.25$ \\     
              &$\Sigma K$     &$- 6.62$  &$     0$ &$- 6.62$ &$- 2.21$ \\     
              &$\Omega K$     &$     0$  &$+ 0.19$ &$- 0.19$ &$- 0.19$ \\     
\noalign{\smallskip}
\hline
\end{tabular}
\end{table}

This approach allows one to separate an individual quark-flavour
contribution into four categories, namely: full, loop, valence, and
quenched valence contributions.  The LNA loop contribution is relevant
to disconnected insertions of the EM current in {\it either} full or
quenched QCD, whereas the LNA valence contribution is relevant to
connected insertions of the EM current only in full QCD.  The final
category of quenched valence contributions is relevant to connected
insertions of the EM current in quenched QCD.  The latter is often
referred to as the quenched QCD result.  

Similar arguments allow one to calculate the contributions of all
quark flavors.  Table~\ref{tab:quarkMom} reports values for the
coefficient, $\chi$, providing the LNA contribution to baryon magnetic
moments ($\chi \, m_\pi$ or $\chi \, m_K$ as appropriate) by quark
sectors normalized to unit charge.  Charge symmetry provides the
contributions for the other baryons and $\Sigma^0$ is the isospin
average of $\Sigma^+$ and $\Sigma^-$.  Quenched coefficients for
$\Lambda$ are determined via SU(3)-flavour symmetry.

Baryon moments are constructed from the quark sector coefficients by
multiplying the $u$, $d$ and $s$ results by their appropriate charge
factors and summing.  For example, the proton moment is $\mu_p = 2 u_p
/ 3 - d_p / 3 - s_p / 3$ and the neutron moment is $\mu_n = - u_p / 3
+ 2 d_p / 3 - s_p / 3$.

\begin{figure}[t]
\centering{\
\rotate{\epsfig{file=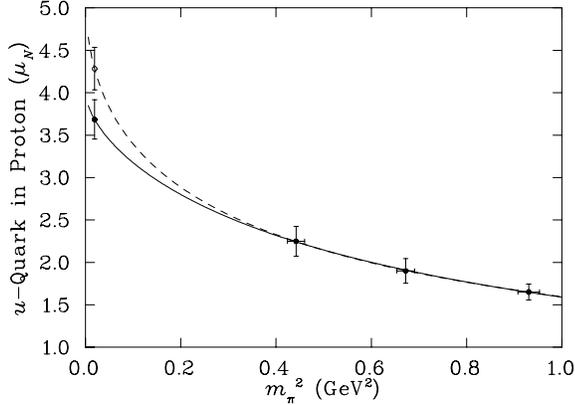,height=\hsize}} }
\vspace{-1.2cm}
\caption{Chiral extrapolation via the Pad\'e of \protect\cite{Leinweber:1998ej}.
The solid curve displays the self-consistent extrapolation of the
quenched simulation results of \protect\cite{Leinweber:1990dv}, whereas the dashed
curve shows the curvature of full QCD.}
\label{uQuarkP}
\end{figure}

These results reveal that the $u$-quark in the proton, known to have
small statistical errors \cite{Leinweber:1990dv}, provides an
excellent opportunity to directly view nonanalytic behavior associated
with the quenched meson cloud of baryons in the quenched
approximation.  Figure~\ref{uQuarkP} illustrates the anticipated
curvature \cite{Leinweber:1998ej} based on the quenched valence
results presented here.

\section*{ACKNOWLEDGEMENTS}

Thanks to Matthias Burkardt, Tony Thomas, Tony Williams and Ross Young
for beneficial discussions.  This research is supported by the
Australian Research Council.


\begin{thebibliography}{9}

\bibitem{Sharpe:1992ft}
S.~R.~Sharpe,
Phys.\ Rev.\ D {\bf 46} (1992) 3146
[hep-lat/9205020],
{\it ibid.}
{\bf 41} (1990) 3233.

\bibitem{Bernard:1992mk}
C.~W.~Bernard and M.~F.~Golterman,
Phys.\ Rev.\ D {\bf 46} (1992) 853
[hep-lat/9204007].

\bibitem{Labrenz:1996jy}
J.~N.~Labrenz and S.~R.~Sharpe,
Phys.\ Rev.\ D {\bf 54} (1996) 4595
[hep-lat/9605034].

\bibitem{Savage:2001dy}
M.~J.~Savage,
nucl-th/0107038.

\bibitem{Dong:1997xr}
S.~J.~Dong, K.~F.~Liu and A.~G.~Williams,
Phys.\ Rev.\ D {\bf 58} (1998) 074504
[hep-ph/9712483].

\bibitem{Wilcox:2000qa}
W.~Wilcox,
Nucl.\ Phys.\ Proc.\ Suppl.\  {\bf 94} (2001) 319
[hep-lat/0010060].

\bibitem{Rijken:1998yy}
T.~A.~Rijken, V.~G.~Stoks and Y.~Yamamoto,
Phys.\ Rev.\ C {\bf 59} (1999) 21
[nucl-th/9807082].

\bibitem{Jenkins:1992pi}
E.~Jenkins, {\it et al.},
Phys.\ Lett.\ B {\bf 302} (1993) 482
[Erratum-ibid.\ B {\bf 388} (1993) 866]
[hep-ph/9212226].

\bibitem{Leinweber:1998ej}
D.~B.~Leinweber, D.~H.~Lu and A.~W.~Thomas,
Phys.\ Rev.\ D {\bf 60} (1999) 034014
[hep-lat/9810005];
E.~J.~Hackett-Jones, D.~B.~Leinweber and A.~W.~Thomas,
Phys.\ Lett.\ B {\bf 489} (2000) 143
[hep-lat/0004006].

\bibitem{Leinweber:1990dv}
D.~B.~Leinweber, R.~M.~Woloshyn and T.~Draper,
Phys.\ Rev.\ D {\bf 43} (1991) 1659.

\end{thebibliography}
\end{document}